\documentclass[10pt, a4paper,english]{article}

\usepackage[utf8]{inputenc}
\usepackage[T1]{fontenc}
\usepackage{babel}
\usepackage{lrec}
\usepackage[hyphens]{url}
\usepackage{array}  % for fixed table width
\newcolumntype{L}[1]{>{\raggedright\let\newline\\\arraybackslash\hspace{0pt}}m{#1}}
\newcolumntype{R}[1]{>{\raggedleft\let\newline\\\arraybackslash\hspace{0pt}}m{#1}}

\title{What do the founders of online communities owe to their users?
}

\name{Cathy Chua$^1$, Manny Rayner$^2$}

\address{$^1$Independent researcher \\
         $^2$Geneva University \\
         cathyc@pioneerbooks.com.au, Emmanuel.Rayner@unige.ch}

\abstract{ %We observe that the ways in which online communities are established offer strong and sometimes irresistible inducements to unethical behaviour on the part of the community's founders. 
We discuss the organisation of internet communities, focusing on what we call the principle of ``bait and switch'': founders of internet communities often find it advantageous to recruit members by promising inducements which are later not honoured. 
We look at some of the dilemmas and ways of attempting to resolve them through two paradigmatic examples, Wikispaces and Wordpress. Our analysis is to a large extent motivated by the demands of CALLector, a university-centred social network we are in the process of establishing. We consider the question of what ethical standards are imposed on universities engaged in this type of activity.\\ \newline }

\begin{document}

\maketitleabstract

%\noindent\textbf{Index Terms}: CALL, reading, hypertext, open source

%\begin{document}

%\maketitleabstract

\section{Introduction and background}

Our point of departure in this paper is CALLector.\footnote{https://www.unige.ch/callector} The overall goal of this new project is to create a social network which will link together producers and consumers of online CALL content; most obviously, this includes teachers, students, content developers who may or may not be teachers, and technical developers. The potential impact of a successful project creates an obligation to organise it in an ethically responsible way. Ethical issues having to do with privacy on the internet have received a great deal of attention over the past few years; for example, they are the topic that receives most attention in the Stanford Encyclopedia of Philosophy article on ``Social Networking and Ethics'' \cite{Vallor2015}. The nature of ``work'' on the internet has also been the subject of some well-cited studies \cite{Terranova2004,BaymBurnett2009,Banks2008}. Certain important aspects of internet work have, however, been comparatively ignored, and it is with some of these that this paper is concerned.

The specific issue on which we will focus is what we will call ``the principle of bait and switch'', which in our experience is regrettably common in the world of online communities. A successful online community has substantial value; a conservative estimate is that each member is `worth' \$10, so a community with millions of members is worth tens of millions of dollars\footnote{An internet search reveals no clear agreement on ways to estimate the value of social networks, with widely differing figures. \$10 per user is near the low end of the spectrum.}. This value largely comes from user input including content and technical development, advice, publicity, or simply time spent on the site which creates atmosphere necessary to a sense of community. At an early stage it is in the interests of the founders to encourage use of the site and development of content by inducements, ranging from perceived prestige to extravagant assurances. A typical motive for this is to then sell the community, often to a large multinational. The users, who might more accurately be called collaborators, have no rights and will not make any money from their unpaid labour and goodwill. If anything, they will find that things start going bad for them.

In the rest of the paper, we start in sections~\ref{Section:Wikispaces} and~\ref{Section:Wordpress} by contrasting two paradigmatic examples, the Wikispaces and Wordpress communities. We include detailed quotes from the creators of these sites, as they have deeply considered some of the issues involved. In section~\ref{Section:UniversitiesAndEthics}, we consider the ethical obligations inherent in university-centred projects like CALLector.  The final section concludes.

%Perhaps the most obvious example of bait and switch on the internet is Facebook. One begins with a startup. Support, publicity and content are crowdsourced. Many fail, but some go on to spectacular success. Addiction is applied, the bait is no longer needed, the switch takes place. Ethical issues in this process to do with privacy have received a great deal of attention over the past few years. Other ethical issues have, however, been all but ignored. They are, in the opinion of the authors, the elephant in the room. It is these with which this paper is concerned.

%CALLector is a Geneva-based project setting up a social network for the development and usage of CALL content and we assume a scenario where it is successful in its aims, supposing that if it fails to get off the ground, there is no case to consider. It envisages various sorts of users and therefore relationships. Most obviously teachers, students, content developers who may or may not be teachers, and technical developers. The over-arching question is how to make the environment achieve ethical outcomes for the users. This then becomes a web of inter-related smaller questions which include:

% \begin{enumerate}

% \item	What are the primary relationships? What are the obligations entailed?
% \item	How to create a sustainable network in terms of numbers of users?
% \item	How to finance it?
% \item	Will content developers be able to sell their products on the site? If so, how will relationships change?
% \item	Ditto for technical developers.

% \end{enumerate}

\section{Wikispaces}
\label{Section:Wikispaces}

%In the first instance we explore these and other questions by examining the case of Wikispaces. Wikispaces (established 2005) is not a perfect match for CALLector as the latter begins from the assumption that it is opensource and non-profit, whereas Wikispaces was initiated as a for-profit business. However, as we shall see, there is much in common and the management of Wikispaces has much to say on the ethics of the relationship between such startups and the education sphere in which they wish to make their mark.
%The Wikispaces community was established in 2005. An early overview, aimed at educators, described it thus:\footnote{http://www.tesl-ej.org/wordpress/issues/volume16/ej62/ej62m2/} 
%\begin{quote}
    %Wikispaces provides wikis for ... tens of millions of users and tens of thousands of institutions ... teachers can create their own wikis for different educational purposes for free. Students may join an existing wiki to participate in and complete class assignments. 
%\end{quote}
The Wikispace community was established in 2005. It allowed teachers to create wikis according to their own requirements, online spaces in which students would then participate. By 2012 it was reported that it had a base of tens of millions of individual users, and many thousands of institutions. 

Of particular significance was the communal collaborative process Wikispaces permitted, with no geographical limitations, as one teacher describes:\footnote{https://www.cnet.com/news/a-quarter-million-teachers-to-get-free-wikis/}
%\footnote{\footurl1}

\begin{quote}
For Vicki Davis, a teacher at the Westwood Schools in Camilla, Ga., the free wikis project has been a boon to developing her students' sense of how to be a responsible online citizen, as well as for completing collaborative projects.

Davis' institution has been part of the Wikispaces project since the beginning, and has engaged in several different online initiatives that have involved more than 1,000 students from public and private schools in many different countries.

She said her students are using the wikis to research the ideas of digital citizenship raised in Thomas Friedman's famous book, The World is Flat: copyright, digital law, digital ethics, and digital etiquette, and are using the wikis to write collaborative reports.

``When they're done (writing), they have a collaborative report and 10 to 15 students from at least six countries have edited it,'' said Davis. ``They learn what it's like to live in a connected world.''
\end{quote}

Multiply this by millions to see the impact that Wikispaces created with their product, linking students and teachers the world over.

In 2012, the directors of Wikispaces made a long statement discussing the ethics of their operation:\footnote{https://www.edsurge.com/news/how-to-succeed-in-ed-tech}

\begin{quote}
    We define success in edtech as \textbf{building a sustainable company that improves student outcomes, empowers teachers, and increases the reach and efficiency of educational institutions.}
\end{quote}

They go on to discuss the moral imperative of sustainability:

\begin{quote}

When an established edtech company fails, it's a big deal. The impact on students, teachers, and administrators is far higher than for similar services outside education. Money for a replacement is tied up in an annual budgeting process. IT and technology support roles--already understaffed--need to juggle this emergency alongside their existing responsibilities. Teachers and administrators simply do not have extra hours during the school year for technology training. Students need to start over with new materials and a new product to learn.
    
These factors mean that when an edtech company closes its doors, their customers are left bearing a heavy burden.

We believe edtech startups have a higher duty--a moral duty--to their students, teachers, and administrators ...
Build products that will survive the test of time. Build companies that will be around to support students and educators beyond the next fad, the next wave of technology change, the next economic downturn. And temper your expectations with a healthy dose of patience. Companies that are built to sustain themselves will be around long enough to find success.

\end{quote}

Although students are at the heart of any such operation --- `Reaching large numbers of students is hard, helping them in a measurable way is harder, and proving that you did is harder still.' --- Wikispaces, unlike many educational startups (in CALL, Memrise\footnote{https://www.memrise.com/} and Duolingo\footnote{https://www.duolingo.com/} come to mind) considers teachers to be of vital importance because:
%\begin{quote}
\begin{itemize}
\item They are the great enablers of student adoption. Teachers decide which products and platforms their classrooms use.
\item They know better than anyone how to help their students succeed. Teachers will show you how to build a better product, but only if you respect their time and the fact that all students, teachers, and schools are different. A great product that requires a 25th hour in the day is not going to get used. A great product that mandates a narrow pedagogy will not achieve broad adoption. When you empower teachers to use technology effectively, it magnifies the impact they can have on their students.
\item 	Teachers exert a large and growing influence on the technology decisions of their institutions. The impact of this final point on ed-tech startups cannot be overstated.’
\end{itemize}
%\end{quote}

CALLector is also teacher-focussed; its expectation is that it is a network for teachers to build a community. Teachers are, with good reason, wary of such sites. It is important to understand and react appropriately to them.

The Wikispaces manifesto next describes some things that success in the field of education isn't. For Fame and Riches, seek more promising arenas. And in particular for us: 

\begin{quote}
Technology innovation in ignorance of customer benefit. Building novel features based on new technology is very satisfying — particularly to engineers — in the short-term. In the long-term, we believe that most innovative products will balance novelty with simplicity, and will always be based on a deep understanding of the customer.
\end{quote}

It urges the model of charging in a fair and transparent way from day one.

\begin{quote}
``Free'' is without question a wonderful marketing tool to get your product in the hands of as many students and teachers as possible. For a company to survive, however, someone must foot the bills. Of the many creative options available, we believe the best source of revenue for an education company is to charge your customers directly for the services they use.
\end{quote}

It is not only companies that must foot bills. However minimal the costs of an open source network are, they are not nothing. The internet bait and switch ploy of free until it isn’t, is unethical at the best of times, but may be catastrophic in the field of education. Nor is advertising revenue an acceptable ethical way to resolve this – if ever, but certainly in regard to education.

In 2014, Wikispaces was sold to TSL for an undisclosed sum. Founders Frey and Byers stated:\footnote{https://www.edsurge.com/news/2014-03-04-tsl-education-acquires-wikispaces}

\begin{quote}
Some of you may be skeptical, thinking that this acquisition may affect our ability to continue to serve teachers as we always have, or that it might change our focus so that we can no longer be the partners to the education community we have prided ourselves on being. To those concerns all we can say is 'watch what happens'.
\end{quote}

Watch what happens? In July 2018, Wikispaces announced its closure.\footnote{http://helpcenter.wikispaces.com/customer/portal/articles/\-2920537-classroom-and-free-wikis} 

\begin{quote}
Free and Classroom Wikis will cease to exist past 31st July 2018 23.59 GMT+1.  After this date, you will be unable to access your data.  Therefore we strongly suggest you take steps to extract any data you wish to retain from the site before this date. 
\end{quote}

The tens of millions of users were given just two weeks to save any data before it was destroyed. 

\begin{quote}
Once Wikispaces has closed the doors for good, your data will be permanently deleted.  Therefore, data will become completely inaccessible to yourself, members, users, the public and our engineers. 
 
As a result of this, I would highly recommend ensuring that you have exported all of your data before the end date of your Wiki to ensure that you have a copy saved.
\end{quote}

It is difficult to imagine how Wikispaces could have been less helpful in assisting the users they were now abandoning:

\begin{quote}
The best option is to export the data from your Wiki and save it to your computer and then use the data to create new pages on a different platform.  Due to the different settings on other websites, we do not have a way of exporting your Wiki direct from Wikispaces to another site. 

Alternatively, you can copy and paste the content direct from your Wiki to your chosen Wiki site.
There are many sites that are similar to Wikispaces and used for education purposes.  We would recommend conducting your own research in order to locate a site right for your needs and set up or contact that company directly.
\end{quote}

They give a few examples of sites that teachers could try and add:

\begin{quote}
Please note that due to team capacity we are unfortunately unable to advise further on alternative sites or assist with the export of your data beyond the information provided here.
\end{quote}

Wikispaces was not only telling teachers they had less two weeks to save their own material, but, even more improbably, to get their students to save theirs. Whatever material might have been saved --- and one suspects much must have been lost --- the community-led collaboration between teachers and groups of learners which extended around the world was destroyed for good. 

We don’t know the story behind this extraordinary rescinding of what they claimed to stand for and what one might reasonably call a betrayal of their tens of millions of users. In selling out to TSL, one of the world’s for-profit education giants, they promised technological improvements for their base. Instead, when they closed they stated:

\begin{quote}
... technology has surpassed the site as more and more Wiki sites became available.  Over the last twelve months we have been carrying out a complete technical review of the infrastructure and software we use to serve Wikispaces users. As part of this review, it has become very apparent that the required investment to bring the infrastructure and code in line with modern standards is very substantial. As such it is no longer financially viable to continue to run Wikispaces long term.
\end{quote}

Was this true? Was it always TSL’s intent to close down Wikispaces? Did it take it over in order to do so? If it is true, why was it done in such a devastating way, ensuring maximum cost to the huge user base? Why did Wikispace even cut all its links dead? Why sell yourself as caring and then act in the most uncaring way possible? 

Furthermore, if it was true, did it matter? Teachers generally don’t want bells and whistles, as Wikispaces knew very well. They want things that work, that are reliable, that are user-friendly. In the educational sphere, we note the example of the group of educational sites known as https://www.anglaisfacile.com/tous.php. It has been running since approximately 2002 and has been recommended to teachers ever since. It is ugly, absolutely set in the past technologically, but this very fact recommends it. It is a straightforwardly free, community based site. 

Open Hub's Project Cost Calculator\footnote{https://www.openhub.net/p/wordpress/estimated\_cost} gives an estimate of the human cost of developing Wordpress, an enormous project which now powers over 30\% of the internet:

Codebase Size: 560,703 lines\\
Estimated Effort: 151 person-years\\
Estimated Cost: \$8,282,611 

At the time of the takeover of  by TSL, it is suggested that Wikispaces' annual revenue was \$20M.\footnote{https://www.crunchbase.com/organization/wikispaces\#section-funding-rounds} And yet Wikispaces said it could not afford to update its software. This despite the fact that Wikispaces' claimed that its specific strategy was to invest in technology rather than extraneous costs such as sales staff. They also said back in 2012:\footnote{https://www.edsurge.com/news/how-to-succeed-in-ed-tech}  `And if you serve a portion of your customers for free, they need to know that they aren't part of a bait-and-switch but that their free usage ultimately contributes to your success.'

We suppose that the founders of Wikispaces made a lot of money when they sold out to TSL. But for the rest of those involved in the development of Wikispaces, and that means every member of the community whose support made that profit-making takeover possible, it was a disaster. How is CALLector to avoid this?

\section{Wordpress}
\label{Section:Wordpress}

Wordpress\footnote{https://en.wikipedia.org/wiki/WordPress} (WP) provides an example of a for-profit business which attempts to balance on the fine line between making money and being ethical. It is the most popular blogging platform in the world (though it has developed  beyond that) and has won many awards for the quality of its open source software and for privacy. We look now at its relationship with its users, its ongoing relationship with its own founding principles and how, therefore, it does on the bait and switch measure.

Like Wikispaces, it has a prominent stress on ethical behaviour. Its Foundation Philosophy states:\footnote{https://wordpressfoundation.org/philosophy/}

\begin{quote}
In order to serve the public good, all of the software and projects we promote should support the following goals:
\begin{enumerate}
\item 	The software should be licensed under the GNU Public License.
\item	The software should be freely available to anyone to use for any purpose, and without permission.
\item	The software should be open to modifications.
\item	Any modifications should be freely distributable at no cost and without permission from its creators.
\item	The software should provide a framework for translation to make it globally accessible to speakers of all languages.
\item	The software should provide a framework for extensions so modifications and enhancements can be made without modifying core code.
\end{enumerate}
\end{quote}

In his own blog, Matt Mullenweg stated in 2010:\footnote{https://ma.tt/2010/09/wordpress-trademark/}

\begin{quote}
Automattic has transferred the WP trademark to the WP Foundation, the non-profit dedicated to promoting and ensuring access to WP and related open source projects in perpetuity. This means that the most central piece of WP’s identity, its name, is now fully independent from any company.

This is really a big deal.

I want to recognize and applaud the courage and foresight of Automattic’s board, investors, and legal counsel who made this possible ... The WP brand has grown immeasurably in the past 5 years and it’s not often you see a for-profit company donate one of their most valuable core assets and give up control. However, I know in my heart that this is the right thing for the entire WP community, and they followed me on that. It wasn’t easy, but things worth doing seldom are ...

Automattic might not always be under my influence, so from the beginning I envisioned a structure where for-profit, non-profit, and not-just-for-profit could coexist and balance each other out. It’s important for me to know that WP will be protected and that the brand will continue to be a beacon of open source freedom regardless of whether any company is as benevolent as Automattic has been thus far. It’s important to me to know that we’ve done the right thing. Hopefully, it’s important to you, too, and you’ll continue your support of WP, the WP Foundation, and Automattic’s products and services. We couldn’t do it without you! 
\end{quote}

The contrast is dramatic. Wikispaces's founders in the end did everything they had argued was unethical. Mullenweg early on safeguarded against the unknown future, takeover, his --- or others' --- human weakness. He did all that could be done to ensure that the ethical principles which initiated Wordpress would be maintained without interference.

In doing so, Mullenweg was not acting only for himself. The ethical desirability of his actions is linked to the core users which make an online community successful in the first place. He avoids the following commonplace pattern. First start with an approach emphasising quality to attract the right sort of people to both form a critical mass and to provide invaluable unpaid development advice. Then, once reputation is established, redefine critical mass, replace quality with quantity because this is where the big money will be. For the core users such fundamental change can be deeply traumatic. If they leave, they may keep their content, but they lose their home, their community. Giving users ownership of their content is ethically necessary, but it is not sufficient. They need control of the community as well.

Despite the ethical philosophy behind the Foundation, over the years since its inception, WP has changed dramatically at a user level. As one may surmise from the name, it was made for words. Now there is pressure to monetise blogs. Pictures have become dominant in the same way as they have internet-wide, and the hosting of those has a price to pay. 

Ad warnings appear on posts telling users to pay for ads to be removed. Chirpy messages tell you to click on somebody else's blog posts because they liked yours. Creating activity for its own sake is a prime motivation of WP now. It’s making money for everybody. Words, as the primary concern, are replaced by clicks. Receive an email advising that somebody has commented on a post and it will include an exhortation to upgrade to a premium model in order to `support your growing audience.'

At the time WP started up, people chose it, above the competition, for a reason. Their goodwill is priceless and without it WP is nothing. But WP has some complicated relationships to cater for ethically. Is the user a commodity or a customer? Does it depend on whether they are a free or premium user? But the best content, which drives people to WP may be from free blogs.

Add to this another aspect which has relevance to CALLector: external support to WP users – WP itself employs a very small number of people – is a revenue generator for an unknown, but very large, number of people and they have a relationship with WP too. What are they? A commodity? A customer? Who is more important to WP? 

Can WP conduct an ethical relationship with both of these? The more complicated and feature-rich it becomes, the more necessary technical support is. Technical support providers outside the official WP fold gain from this. 

So far WP has been fairly good at not falling off the tightrope. We hope that CALLector, which shares many of these potential conflicts and dilemmas, will be a better model again.

\section{Ethical obligations of university-based projects}
\label{Section:UniversitiesAndEthics}

The examples of Wikispaces and Wordpress, as well as any number of others specifically in the CALL domain in which the CALLector project is operating, are startups established with a view to making a profit. Some of those, for example Babbel\footnote{https://www.babbel.com/} and RosettaStone\footnote{https://www.rosettastone.co.uk/} are overtly run as businesses with straightforward relationships. They provide a service for money. Others, like Duolingo and Memrise, have a more complicated relationship with the user. They provide something ‘free’ but of course there is always a price to pay. Memrise users are discovering that at the moment, as the startup owners make massive changes to the way in which the site is now run.\footnote{https://community.memrise.com/t/important-update-upcoming-changes-to-memrise-community-created-courses/33461/17} CALLector, in contrast, is a project based at Geneva University with funding from a noncommercial source, the Swiss National Science Foundation, which states on its site:\footnote{http://www.snf.ch/SiteCollectionDocuments/snf\_leitbild\_e.pdf}

\begin{quote}
    ‘Our commitment to the public: Our work promotes the spread of knowledge in society. We ensure access to research results and communicate them to the public. We show how research contributes to social progress, economic growth and a high quality of life.’ 
\end{quote}

Around the world the idea of University Social Responsibility, a spinoff from Corporate Social Responsibility, is to be seen governing institutions of higher education (http://www.usrnetwork.org/; \cite{VasilescuEA2010}). The USR Network, for example, explains the rationale for its establishment thus: ‘Based on the belief that universities have obligation to work together to address the economic, social, cultural and environmental challenges in the world and to find solutions so as to make our world more just, inclusive, peaceful and sustainable…’ This idea has been part of the EU’s higher education for a long time \cite{VasilescuEA2010,SchnellerEA2011,WallaceResch2017}.

Emanating from the US is HASTAC which is\footnote{https://www.hastac.org/blogs/superadmin/2011/08/16/hastac-defined-and-numbers} 

\begin{quote}
‘a network of individuals and institutions inspired by the possibilities that new technologies offer for shaping how we learn, teach, communicate, create, and organize our local and global communities.  We are motivated by the conviction that the digital era provides rich opportunities for informal and formal learning and for collaborative, networked research that extends across traditional disciplines, across the boundaries of academe and community, across the "two cultures" of humanism and technology, across the divide of thinking versus making, and across social strata and national borders. Participation is our leadership model and collaboration by difference is our guiding method. HASTAC's mission is shaped by the active participation and interests of our members. We are what our members make us. As a "virtual organization" whose work centers on weaving together people and ideas from across disciplines, HASTAC's web site is both a platform for convergence and a stage for experimentation and practice.’ 
\end{quote}

It is evident from the literature that however obvious the idea of ‘social responsibility’ is, defining it is not so clear. Obligations to address challenges, solutions for a better world, are problematic in fruition \cite{Weiss2016} not least due to issues of funding \cite{Shek2017usr}. Recent minutes for the EU’s Advisory Group on the Social Dimension of Higher Education\footnote{http://www.ehea.info/Upload/AG1\_SD\_1\_Minutes.pdf} discuss some of the issues involved in bringing greater equality to higher education within Europe  and if this is an issue, one can surmise that the broader remit of obligations to society at large will not be easily fulfilled. 

Against this background and despite the difficulties, however, one can see that universities are well placed to provide an ethical open-source resource of the type required by the CALLector project; the principles of USR also suggest that it has the obligation to do so. To arm educators at school level with appropriate technology for better learning will ultimately advantage not only the schools and their students, but also universities who will reap a reward from students who come better educated in general. That is to say, however altruistic it may appear to provide these resources and develop the social network framework for their best use, in the end higher education can expect a payout. It’s an investment, not a gift.

Various quotes from \cite{SchnellerEA2011}\footnote{All taken from  Isabelle Turmaine and Chripa Schneller, ``Universities' contribution to Education for All (EFA) and the Millennium Development Goals (MDGs)'' http://www.seaairweb.info/Collaborations/2011USR\_ASEF.pdf} reinforce this notion: 

\begin{quote}
`... no country can build an effective higher education sector without human resources and quality basic and secondary education. Inevitably higher education and research should also be involved –-- as part of its social responsibilities –-- in the promotion of other education levels.' 

`... it is the responsibility of universities to eliminate barriers to higher education and integrate non-traditional students, thus to ensure alternative pathways of access.'

`In today’s global, fast changing, but also critical world, universities need to be aware that they serve the society at large more than ever before. Therefore, they need to revisit their role, assume social responsibility as an evidence-based concept and foster sustainable development.' 
%Their mission cannot be built only on an academic base anymore. Higher education policy should consequently not be detached from social policy in order to secure a promising, just and environmentally sustainable future for our societies. As USR does involve investments and therefore costs, governments need to secure funding for the further development of USR, which encompasses wider aspects than Corporate Social Responsibility (CSR), such as international links in teaching, research and services.’

‘Universities should particularly be supported in communicating and exchanging good and innovative ideas with the general public.’ 

‘... universities in ASEM countries should reflect on the entire education process, from early childhood education to lifelong learning.’ 
\end{quote}

\section{Conclusion}

Exploiting the addictive potential of the internet is a business model. Offer free/generous terms, and then, once dependency or addiction has set in, make 'em pay. Monetisation is the name of the game. The business argument is that ethics don't come into it. They need only obey the letter of the law, or exploit its greyness. 

Universities, however, are not businesses. Their raison d’être is not to make money. They have a relationship with, and obligation to, the community. The question to be asked, therefore, is can CALLector, a university initiated project, avoid these ethical dilemmas? 

\section{Bibliographical References}

\bibliographystyle{lrec}
\bibliography{proposal}

\end{document}